\documentclass[10pt,noshowpacs,amsmath,twocolumn,
aps,prb]
{revtex4-1}

\usepackage{setspace}
\usepackage{amsmath}
\usepackage{graphicx}
\usepackage[nearskip,margin = 0pt]{subfig}
\usepackage{subfig}
\usepackage{verbatim}
\usepackage{amsfonts}
\usepackage{amssymb}
\usepackage{epstopdf} 
\usepackage{ulem}
\usepackage{xcolor}
\DeclareGraphicsExtensions{.pdf,.eps,.png,.jpg,.mps}
\usepackage{ragged2e}
\usepackage{hyperref}
\hypersetup{
    colorlinks=true,
    linkcolor=blue,
    citecolor=blue,
    filecolor=blue,      
    urlcolor=blue,
}

\begin{document}

\title{Efficient and wavelength-tunable second-harmonic generation towards the green gap}

\author{Zhiquan Yuan$^{1,*}$, Jinhao Ge$^{1,*}$, Peng Liu$^{1,*}$, Bohan Li$^{1,*}$, Mingxiao Li$^{2}$, Jin-Yu Liu$^{1}$, Yan Yu$^{1}$, Hao-Jing Chen$^{1}$, John Bowers$^{2}$, and Kerry Vahala$^{1,\dagger}$\\
$^1$T. J. Watson Laboratory of Applied Physics, California Institute of Technology, Pasadena, CA 91125, USA.\\
$^2$ECE Department, University of California Santa Barbara, Santa Barbara, CA 93106, USA.\\
$^*$These authors contributed equally to this work. \\ 
$^\dagger$Corresponding author: vahala@caltech.edu}

\maketitle

{\bf \noindent Abstract} 

Achieving compact and efficient visible laser sources is crucial for a wide range of applications. However traditional semiconductor laser technology faces difficulties in producing high-brightness green light, leaving a “green gap” in wavelength coverage. Second-harmonic generation (SHG) offers a promising alternative by converting near-infrared sources to visible wavelengths with high efficiency and spectral purity. Here, we demonstrate efficient and tunable SHG within the green spectrum using a high-$Q$ Si$_3$N$_4$ microresonator. A space-charge grating induced by the photogalvanic effect realizes reconfigurable grating numbers and flexible wavelength tuning. Additionally, grating formation dynamics and competition is observed. These findings underscore the potential of silicon nitride as a robust, integrative platform for on-chip, tunable green light sources.

\section{Introduction}
The development of compact, coherent visible laser sources is
key to a range of applications in both science and engineering, including optical clocks \cite{ludlow2015optical}, biomedical imaging \cite{luke2019lasers}, quantum information processing \cite{wendin2017quantum}, and laser displays \cite{chellappan2010laser}.
However, it is challenging to create green light with high brightness and efficiency from traditional semiconductor laser technologies due to material limitations and low quantum efficiency \cite{moustakas2017optoelectronic}. 
This issue is commonly referred to as the “green gap” \cite{pleasants2013overcoming}, a wavelength span around 500 – 550 nm.
Dye lasers based on organic molecules offer wide wavelength tunability. However, like other traditional bench-top solutions, they are bulky and complex, limiting their use in compact applications \cite{schafer2013dye}. 
Also, quantum dot lasers have high tunable range and are suitable for on-chip integration, but still suffer from low efficiency and coherence at the green gap \cite{mei2017quantum, zhao2020low}.

Nonlinear optical phenomena in high-Q resonators provide another approach for achieving new wavelengths \cite{vahala2003optical}. For instance, optical parametric oscillation (OPO) based on four-wave mixing (FMW) in microresonators \cite{kippenberg2004kerr, savchenkov2004low} has been extensively studied in recent years \cite{sayson2019octave, lu2020chip, domeneguetti2021parametric}, and enables access to the green gap \cite{sun2024advancing}. Alternatively, green light generation has been reported by third-harmonic generation \cite{carmon2007visible, levy2011harmonic, surya2018efficient} and cascaded sum frequency generation \cite{ling2022third}, but with limited efficiency.
In contrast, second-harmonic generation (SHG), offers high conversion efficiency and coherence by frequency-doubling from near-infrared sources \cite{lu2019periodically, lu2020toward, ling2023self}.
Among various photonic platforms for SHG, Si$_3$N$_4$ is particularly promising due to its ultra-low optical loss, CMOS compatibility and suitability for nonlinear optics. 
And recently, photogalvanic effect induced second order nonlinearity has been observed in Si$_3$N$_4$ with efficient SHG \cite{puckett2016observation, billat2017large, lu2021efficient, nitiss2022optically, li2023high}. 

\begin{figure*}[t!]
\begin{centering}
\includegraphics[width=\linewidth]{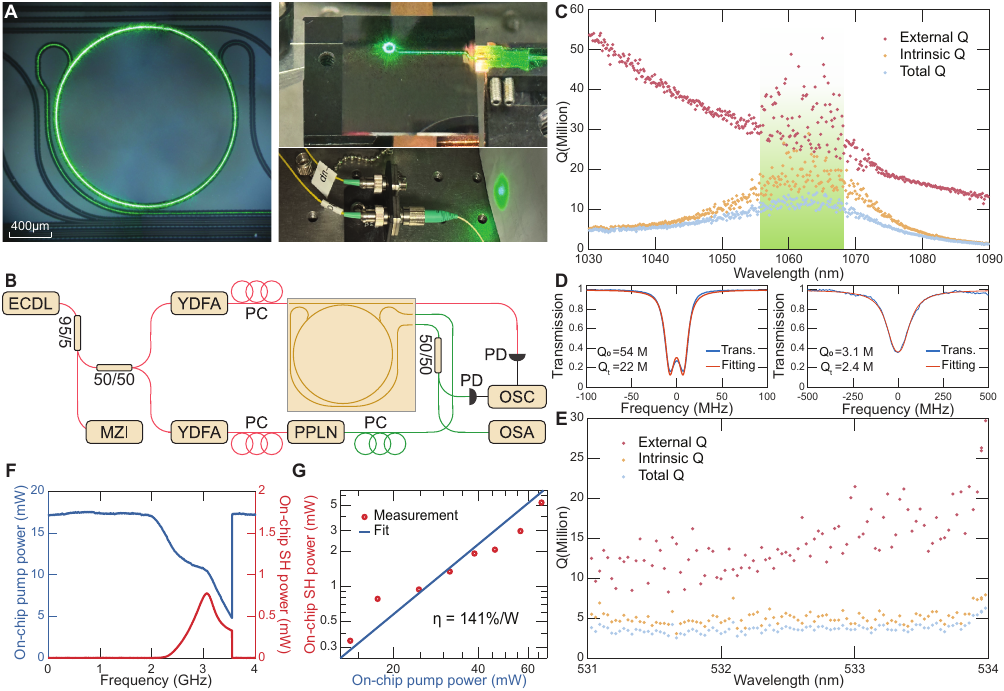}
\captionsetup{singlelinecheck=off, justification = RaggedRight}
\caption{{\bf Device characterization and SHG performance of the Si$_3$N$_4$ microresonator.}
{\bf a} Photographs of the ring resonator and off-chip green light emission during SHG operation.
{\bf b} Experimental setup for frequency-matching tuning and SHG performance characterization of the Si$_3$N$_4$ microresonator. 
The chip temperature is controlled and stabilized by a thermoelectric cooler (not shown).
Abbreviations: ECDL, external cavity diode lasers; MZI, Mach–Zehnder interferometer; YDFA, Ytterbium-doped fiber amplifier; PPLN, periodically poled lithium niobate; PC, polarization controller; PD, photodetector; OSC, oscilloscope; OSA, optical spectrum analyzer.
{\bf c} Measured quality ($Q$) factor distribution versus wavelength from 1030 nm to 1090 nm, with the designed high-$Q$ band highlighted in green.
{\bf d} Transmission spectrum and fitting of the pump resonance at 1064 and 532 nm.
{\bf e} Measured $Q$ factor distribution versus the wavelength from 531 nm to 534 nm. 
{\bf f} On-chip transmission of pump power (left axis) and SHG power (right axis) during pump laser frequency scanning across a cavity resonance at the phase-matching condition.
{\bf g} On-chip SHG power (red dots) versus pump power. The frequency conversion efficiency is fit by the blue line with a slope of two, indicating a constant SHG efficiency ($\eta$, in \%/W) within the measured range.
}
\label{Fig1}
\end{centering}
\end{figure*}

In this work, we leverage a thin-film, ultra-low-loss Si$_3$N$_4$ \cite{jin2021hertz} to achieve efficient SHG in the green spectral range. By employing a carefully designed coupler to minimize leakage loss, we demonstrate a broadband high-$Q$ region and strong green emission with high coherence (see Fig. \ref{Fig1}a). 
In the photogalvanic-induced quasi-phase-matching process, optically induced space-charge gratings are reconfigurable, enabling robust, dynamically tunable SHG in real time. Furthermore, we observe dynamic grating competition during space-charge formation. The tunability, combined with the maturity of silicon nitride as a platform for photonic integration, positions this system as an ideal candidate for versatile, on-chip green laser sources.

\begin{figure*}[t!]
\begin{centering}
\includegraphics[width=\linewidth]{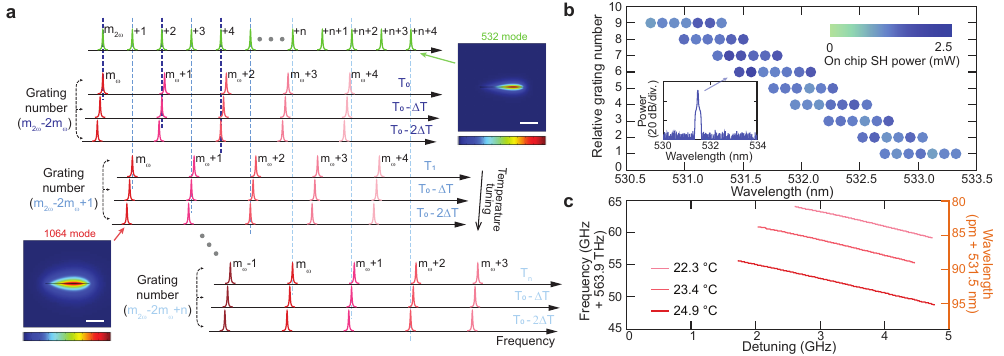}
\captionsetup{singlelinecheck=off, justification = RaggedRight}
\caption{{\bf Coarse and fine wavelength tuning in green SHG.} 
{\bf a} Illustration of the multiple frequency matching condition in the SHG process for coarse tuning control. The first row shows multiple longitudinal modes of the fundamental TE mode family in the green spectral region. 
Subsequent rows display the longitudinal modes of the fundamental TE mode at 1.06 $\mu$m, with different SHG frequency-matching conditions (dashed lines) at the corresponding cavity temperatures.
Within each group, the grating number ($m_{\text{green}} - 2m_{\text{near-IR}}$) of the photogalvanic space-charge field remains constant.
Inset: cross-sectional view of simulated mode profiles at the green SHG mode and near-IR pump mode, respectively. The scale bar represents 2 $\mu$m.
{\bf b} Experimental measurements of the generated SHG wavelengths at different matching conditions. 
The y-axis represents the grating number difference between the pump and SHG modes. The dots are shaded to indicate the corresponding on-chip SHG power.
Inset: a typical optical spectrum of the generated green light.
{\bf c} Measured output green light frequencies are plotted versus the laser-cavity detuning at three different cavity temperatures for the same pump-SHG mode pair. The on-chip pump power is 78 mW.
}
\label{Fig2}
\end{centering}
\end{figure*}

\section{Device characterization}
The device is fabricated based on an ultra-low-loss Si$_3$N$_4$ platform \cite{jin2021hertz, puckett2021422}. The Si$_3$N$_4$ resonator waveguide core is a 5-$\mu$m wide by 50-nm thick design with an upper-cladding of 2.2-$\mu$m thick silica. The design has a radius of 850 $\mu$m, corresponding to a free-spectral-range (FSR) of 36.61 GHz at 1064 nm. To facilitate efficient coupling in both the near-infrared (near-IR) and visible bands, the design incorporates two pulley couplers (see Fig. \ref{Fig1}a) for spectral partitioning. The upper coupler, with a 3.5 $\mu$m gap, enables injection of the near-IR pump laser while minimizing coupling to the visible (green) mode, and the lower coupler, with a 0.3 $\mu$m gap, efficiently extracts the SHG signal while reducing the leakage of the near-IR mode.

The microresonator is characterized using the setup illustrated in Fig. \ref{Fig1}b. A 1064 nm tunable laser is split into two beams. One beam serves as the probe wave for the fundamental near-IR modes and provides the pump power for SHG, while the other beam is amplified and frequency-doubled by a periodically poled lithium niobate (PPLN) crystal to probe the green resonances. 
A four-channel V-groove array (VGA) and a lensed fiber are used for simultaneous coupling to both the near-infrared and green waveguides.

The quality factor distribution in the near-IR band is first measured using a calibrated Mach–Zehnder interferometer (MZI) in combination with a wavelength-tunable laser\cite{yi2015soliton}. Due to the wavelength coverage limitation of the PPLN crystal, the center of the low leakage region is designed for 1064 nm. As shown in Fig. \ref{Fig1}c, the measured $Q$ distribution aligns well with this design target, with minimal leakage loss (highest $Q$ factor) centered near 1063 nm. Within 12 nm bandwidth, the total $Q$ factor exceeds 10 million (highlighted in green), ensuring efficient SHG over a wide spectral coverage. The scatter in the $Q$ data in this region is an artifact of measurement near the critical coupling point.
Additionally, we characterize the $Q$ distribution in the green modes by scanning the pump frequency while simultaneous tuning the PPLN crystal temperature to maintain sufficient SHG probe power (Fig. \ref{Fig1}e). Within the measurement range, the total $Q$ factors of the green modes are clustered around 3 million, which shows the low absorption loss of the material \cite{corato2024absorption,morin2021cmos}.

In the photogalvanic process, a space-charge grating creates a field that works with the existing Kerr effect to create an effective second-order nonlinearity. The grating also ensures quasi-phase-matching by compensating for the phase difference between fundamental and SHG fields \cite{nitiss2022optically}. Therefore, only the frequency matching condition is necessary, where the frequency of the visible mode is required to be twice of the fundamental field. Due to the configuration of the PPLN doubler, a green mode needs to align with a near-IR mode during frequency scanning. To accomplish this, chip temperature is tuned to align green and near-IR modes. During this process, the green resonance is monitored by PPLN doubling of the pump laser. At a chip temperature of 25.0 $^\circ$C, a pump resonance (1063.390 nm) aligns with a visible resonance (531.695 nm), and photogalvanic-induced SHG occurs automatically. The $Q$ factors of the pump and SHG modes are shown in Fig. \ref{Fig1}d. When 17 mW input pump laser is launched into the waveguide and scanned across the near-infrared resonance, SHG power as high as 0.78 mW is generated on chip, corresponding to an efficiency of 250 \%/W (Fig. \ref{Fig1}f).
For the same resonance, the on-chip SHG power is also measured versus pump power in Fig. \ref{Fig1}g, with a conversion efficiency of 141 \%/W averaged over input power levels. The SHG power reaches up to 5.3 mW in the bus waveguide at the input power of 67 mW.

\begin{figure*}[t!]
\begin{centering}
\includegraphics[width=\linewidth]{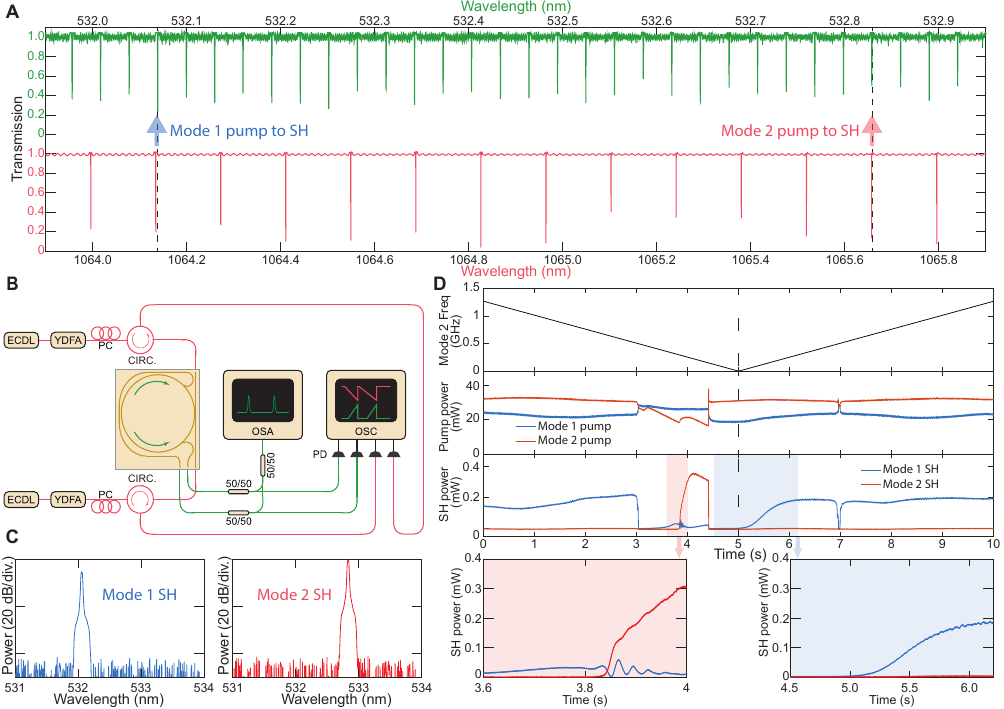}
\captionsetup{singlelinecheck=off, justification = RaggedRight}
\caption{{\bf Vernier matching between green and near-IR modes and observation of grating competition dynamics.}
\textbf{a} Simultaneously measured frequency dispersion of the mode family in both green (green) and near-IR (red) bands is plotted versus wavelength at chip temperature 43.8 $^\circ$C. A vernier frequency matching is observed due to spectrally local FSR differences.
Red and blue arrows indicate the positions of two distinct modes used for SHG.
\textbf{b} Experimental setup for counter-pumped SHG at two distinct wavelengths (mode 1 and mode 2 in panel \textbf{a}).
\textbf{c} Optical spectrum of the independently generated green light from two distinct pairs of modes.
\textbf{d} Observation of the grating competition dynamics between two pairs of modes with different space charge grating numbers. The pump frequency of mode 2 is scanned periodically while the pump frequency of mode 1 is unchanged and in resonance.
Top trace: the frequency change of the mode 2 pump laser during scanning; middle trace: the transmitted on-chip pump power during scanning; 
bottom trace: the corresponding generated on-chip SHG power.
Lower panel insets: zoom-in views of the corresponding shaded regions of the bottom trace, showing the grating built-up and erasing process.
}
\label{Fig3}
\end{centering}
\end{figure*}

\section{Wavelength tunability}
Concerning tuning of the emission, suppose at chip temperature $T_0$, a one-to-one frequency matching condition is satisfied, where a SHG mode $m_{2\omega}$ is frequency matched with the pump mode $m_{\omega}$, $f(m_{2\omega}) = 2f(m_{\omega})$ (as shown in Fig. \ref{Fig2}a). 
Since the green and near-IR modes have different local FSRs and resonant frequency tuning coefficients ($\delta f / \delta T$) near these two modes, the frequency mismatch between the next pair of modes ($m_{2\omega}+2$ and $m_{\omega}+1$) can be compensated through a reduction in chip temperature by $\Delta T = \frac{2\text{FSR}(m_{2\omega}) - 2\text{FSR}(m_{\omega})}{\delta f / \delta T (m_{2\omega}) - 2\delta f / \delta T (m_{\omega})}$. 
This process can be cascaded to achieve multiple frequency-matched output wavelengths, with the optical-poling-induced grating number ($m_{\text{green}} - 2m_{\text{near-IR}}$) remaining unchanged.
The tuning mechanism is illustrated by multiple frequency matching conditions within a group of the same grating number (Fig. \ref{Fig2}a), with each group corresponding to a single row in panel b. The tunability is limited by the chip temperature range, which is kept between 17 $^\circ$C and 60 $^\circ$C, allowing a 0.6 nm wavelength change in the green by steps of 0.1 nm.

Unlike non-centrosymmetric materials poled by high-voltage electrical pulses, where the resonator grating number is fixed during fabrication, the photogalvanic process allows for reconfigurable grating structures due to optically-induced space-charge distribution \cite{nitiss2022optically}. 
Accordingly, and in contrast to the above tuning process, the pump mode $m_{\omega}$ can be matched to adjacent green modes (temperature tuning), such as $m_{2\omega}$ + 1 (second group in Fig. \ref{Fig2}a) by adjusting the temperature. In this case, $T_1 - T_0 = \frac{\text{FSR}(m_{2\omega})}{2\delta f / \delta T (m_{\omega}) - \delta f / \delta T (m_{2\omega})}$. 
This process enables matching a single pump mode to multiple green modes, creating new rows in Fig. \ref{Fig2}b. All experimental measured SHG conditions are summarized and sorted by the grating number, with the on-chip pump power maintained at 39 mW. The OSA spectrum (inset) of each generated green light is used to determine the SHG wavelength and on-chip SHG power. The wide high-$Q$ distribution at both wavelengths enables a maximum tunability of 2.6 nm (2.6 THz) with nearly constant SHG efficiency and minimum on-chip SHG power over 1 mW. 
Importantly, there is no fundamental limit to the SHG wavelength tuning range, and the only restriction comes from the PPLN crystal in the probe of mode resonance frequencies.

Finally we also demonstrate the continuous fine tuning of the SHG wavelength by adjusting the pump laser frequency (laser-cavity detuning) within a single matching point in panel b. Due to the OSA's resolution limit, a wavemeter [HighFinesse GmbH, WS6-600] is utilized to record the green light with 0.1 pm resolution. As shown in Fig. \ref{Fig2}c, at 24.9 $^\circ$C the SHG signal emerges when the laser-cavity detuning reaches 1.8 GHz.
The SHG frequency then decreases at twice the rate of the detuning as it continues to increase, with the laser sweeping out of the cavity mode when the detuning reaches 4.8 GHz. 
When the chip temperature is slightly adjusted while maintaining the frequency matching condition, the cavity resonances also shift,  thereby shifting the green light wavelength. By selecting three cavity temperatures, the output frequency can be continuously tuned by 16 GHz, which means each data point in Fig. \ref{Fig2}b can occupy a width of 16 pm. 

\section{Vernier matching and grating competition}
Due to different local FSRs of modes at 532 nm (32.15 GHz) and 1064 nm (36.61 GHz), a Vernier frequency matching condition can be achieved.
This allows alignment of mode pairs with different grating numbers at the same temperature.
Following the method used in Fig. \ref{Fig1}c and \ref{Fig1}e, we measured the frequency dispersion of the modes in both the green and near-IR spectral bands simultaneously (shown in Fig. \ref{Fig3}a).
At a resonator temperature of 43.8 $^\circ$C, two distinct SHG alignments are observed at 1064.135 nm / 532.069 nm and 1065.658 nm / 532.829 nm. These are indicated by the blue (mode 1) and red (mode 2) arrows, respectively.
The aligned modes are separated by 25 (11) modes in the green (near-IR) band, consistent with the FSR values previously measured (32.15 / 36.61 $\approx$ 22 / 25).

To distinguish the green light generated by each of the two modes, a counter-propagating (CP) pump setup is employed (Fig. \ref{Fig3}b). Here, two pump lasers are amplified and directed through circulators, generating two on-chip counter-propagating SHG signals. These signals are detected independently using photodetectors connected to the two output ports of the VGA. When a single pump laser is coupled into the resonator, SHG occurs automatically upon adjusting the pump wavelengths to match both resonances. The output green light spectra are shown in Fig. \ref{Fig3}c. At this specific temperature, mode 2 achieves perfect frequency matching, while mode 1 has a slight residual misalignment (see panel a), resulting in higher output power for mode 2.

The SHG process arises from photogalvanic-induced grating poling, where the SHG signal follows the formation of the corresponding grating in the microresonator. However, since these two green light signals originate from the same transverse mode family but differ in grating numbers, they produce unique electron distributions within the material. And this introduces grating competition during the formation process of each grating.

To investigate the dynamics of this grating competition, we apply a 0.1 Hz ramp signal to gradually scan the pump laser frequency relative to mode 2, while keeping the mode 1 pump laser in resonance and fixed.
In the first 5 seconds, the mode 2 pump laser frequency gradually shifts from blue to red, scanning from outside of mode 2, entering its resonance, and then moving out again. During the next 5 seconds, it reverses direction, scanning from red to blue across the mode 2 resonance. This frequency scan path is depicted in the top trace of Fig. \ref{Fig3}d, with the corresponding transmitted pump power and SHG power on-chip displayed in the middle and bottom traces. 
At the beginning, the mode 1 pump laser is tuned into mode 1 resonance, establishing the grating structure for mode 1 and enabling SHG. As the mode 2 pump frequency gradually scans into resonance, the SHG power from mode 2 increases while the SHG power from mode 1 decreases (zoom-in in the left inset).
This observation indicates the formation of the grating for mode 2 and the erasure of the mode 1 grating, leading to a brief overlap where both green lights coexist. Fluctuations in mode 1 SHG power are attributed to interference induced by backscattering from mode 2. The grating formation for mode 2 occurs within a second, consistent with previous observations \cite{li2023high}.

When the mode 2 pump laser scans out of resonance, it no longer influences the grating structure, allowing the mode 1 pump laser to re-establish resonance and rebuild its grating. Since the grating for mode 2 has already been formed, mode 1 SHG power does not regenerate immediately but instead increases gradually as the mode 1 grating is rebuilt and mode 2’s grating neutralizes (right inset of Fig. \ref{Fig3}d). The observed relaxation time between the formation of the two gratings reflects the timescale associated with dynamic switching between modes. mode 1’s grating takes longer to regenerate due to the slight mode misalignment.

For comparison, at $t = 7$ seconds, mode 1 SHG power drops as the mode 2 pump laser scans into resonance from the red side and change the cavity temperature. However, this brief resonance does not last long enough to create a new grating or fully erase the existing one, so mode 1 SHG power quickly returns after the mode 2 laser exits resonance.

\section{Discussion}
In summary, we have demonstrated efficient and tunable second-harmonic generation in silicon nitride microresonators within the green spectral range. By leveraging photogalvanic induced quasi-phase-matching, we achieved robust second harmonic generation with precise control over multiple frequency-matched modes. 
Detailed quality factor characterization for both green and near-IR modes highlights the potential of Si$_3$N$_4$ microresonators for high-efficiency, stable performance across different frequency bands.
The observed grating competition between SHG modes further reveals the dynamic behavior of photonic grating formation, suggesting avenues for more advanced control, including dynamic mode-switching. 

The results presented here not only show the versatility of silicon nitride as a low-loss platform for nonlinear optics but also point to its potential for broader photonic integration, including active components such as III-IV semiconductor lasers \cite{komljenovic2015heterogeneous, xiang2021laser, moss2013new, dutt2015chip, stern2018battery, lu2019chip, snigirev2023ultrafast}. 
With high efficiency and compatibility with on-chip integration, Si$_3$N$_4$ microresonators emerge as promising candidates for integrated, tunable on-chip green laser sources.

\vspace{3 mm}

\noindent \textbf{Data Availability}
The data that supports the plots within this paper and other findings of this study are available from the corresponding author upon reasonable request.

\vspace{1 mm}

\noindent \textbf{Code Availability}
The codes that support findings of this study are available from the corresponding author upon reasonable request.

\vspace{1 mm}
\noindent \textbf{Acknowledgments}
The authors thank Tower Semicondutor for sample fabrication. This work is supported by the Defense Advanced Research Projects Agency (HR001-20-2-0044), the Air Force Office of Scientific Research (FA9550-23-1-0587) and the Kavli Nanoscience Institute at Caltech.
\vspace{1 mm}

\noindent\textbf{Author Contributions} Measurements and modeling were performed by Z.Y. and J.G., with the help from B.L., J.-Y.L., Y.Y. and H.-J.C. Structures were designed and prepared by P.L. and M.L. All authors contributed to the writing of the manuscript. The project was supervised by J.B. and K.V.
\vspace{1 mm}

\noindent \textbf{Competing Interests} The authors declare no competing interests.

\vspace{1 mm}

\noindent \textbf{Author Information} Correspondence and requests for materials should be addressed to KV (vahala@caltech.edu).


\end{document}